\documentclass{jpsj3}
\usepackage{txfonts}

\def\gs{> \kern -12pt \lower 5pt \hbox{$\displaystyle{\sim}$}}
\def\ls{< \kern -12pt \lower 5pt \hbox{$\displaystyle{\sim}$}}

\def\ve{\varepsilon} 
\def\be{\begin{equation}}
\def\bea{\begin{eqnarray}}
\def\ena{\end{eqnarray}}
\def\en{\end{equation}}
\newcommand{\bi}[1]{\mbox{\boldmath$#1$}}
\newcommand{\av}[1]{\langle{#1}\rangle}
\newcommand{\tensor}[1]{\stackrel{\leftrightarrow}{{#1}}}

\def\p{\partial} 

\title{Selective solvation in aqueous 
mixtures: Interface deformations 
and instability }

\author{Akira Onuki and Takeaki Araki 
}
\inst{Department of Physics, 
Kyoto University, Kyoto 606-8502, Japan}
\abst{We briefly review 
the effects of selective solvation 
of ions in aqueous mixtures, 
where the ion densities and   the composition 
fluctuations are  strongly coupled. 
We then  examine the surface tension $\gamma$ 
of a liquid-liquid interface in the presence of ions. 
We show that $\gamma$  can be 
decreased  drastically due to the electrostatic 
 and solvation interactions  
near the interface.  
We calculate how the free energy is changed due to small 
surface undulations in the presence of an 
electric double layer. A  surface instability 
occurs for negative $\gamma$, which can easily 
be realized for antagonistic ion pairs near the solvent 
criticality. 
Three-dimensional simulation shows 
how the  surface instability is induced. 
}

\kword{Selective solvation, salt effect, liquid-liquid interface,
antagonistic salt}

\begin{document}
\maketitle

\section{Introduction}

In usual electrolyte theories,  
ions interact via the Coulombic 
potential in a fluid with a homogeneous 
dielectric constant $\ve$. However,  
in  most of such theories, 
the microscopic  molecular interactions 
between  ions  and  solvent 
 molecules   are not explicitly 
considered  \cite{Is,Araki,Nara,Current}. 
Around a microscopic ion such as Na$^+$ or Cl$^-$ 
in a polar fluid,  the ion-dipole interaction 
gives rise to a solvation (hydration) shell composed 
of a number of   solvent molecules 
(those of the  more polar component 
for a mixture). 
The resultant solvation free energy per ion 
will  be called the solvation 
chemical potential and 
will be written as $\mu_{\rm sol}^i$  
where $i$ represents the ion species. 
It is important that 
   $\mu_{\rm sol}^i$ 
strongly depends  on the solvent  density and/or 
the composition for binary mixtures, with 
its typical values much larger than 
 the thermal energy $k_{B}T$.   
In the supercritical region of  pure water 
the density-dependence of    $\mu_{\rm sol}^i$ 
is very large. 
For binary mixtures \cite{Onuki-Kitamura,OnukiPRE}, 
let us introduce its 
derivative with respect to the composition $\phi$:  
\be 
g_i= - (k_BT)^{-1}\p \mu_{\rm sol}^i/\p \phi, 
\en 
which represents the degree of selective or preferential 
solvation of ions. 
Hereafter,  $\phi$ is the water 
composition in water-oil.  
Then  $g_i>0$ for hydrophilic ions 
and $g_i<0$ for hydrophobic ions. 
For strongly hydrophilic ions, 
 $\mu_{\rm sol}^i(\phi)$ should decrease very steeply 
  in a concentration  range 
 $0<\phi<\phi^i_{\rm sol}$, where 
 hydration shells are formed  \cite{Current}. 
 Here  $\phi^i_{\rm sol}
 \sim 0.002$ for Br$^-$ in water-nitrobenzene. 
In this paper, we discuss the selective solvation effects 
in the range  $\phi>\phi^i_{\rm sol}$ assuming 
the presence of well-defined hydration shells 
even in the water-poor region, where 
the variation of 
$ \mu_{\rm sol}^i(\phi)$ is    milder 
but $g_i$ is still large (being of order  
10 for monovalent ions) \cite{Current}. 
Furthermore, for 
polyelectrolytes and colloids, dissociation of ions 
can occur at their ionizable groups and 
the degree of dissociation  sensitively 
depends on the local composition \cite{Bu2,Onuki-Okamoto1,Onuki-Okamoto2}.
Therefore,  the ion densities and 
the solvent composition are strongly coupled 
ubiquitously in soft matter. 
Around colloids, polymers,  proteins, 
and so on,   very small 
 heterogeneities in the solvent composition  drastically alter  
  the ion distributions and the local electric potential.  
As  our recent findings, 
this  selective solvation induces  
(i) prewetting transitions on the surfaces of 
 ionizable rods (polyelectrolytes)\cite{Onuki-Okamoto2} 
 and spheres (colloids)\cite{Okamoto-colloid}, 
(ii)  phase separation (precipitation) induced by 
a small amount of highly selective solute 
 even outside the solvent coexistence 
curve \cite{Okamoto,Opinion},  
and (iii) mesophase formation 
in aqueous mixtures for an antagonistic salt  
\cite{Araki,Nara,Current,Seto1,Seto2,Seto3}.

In aqueous systems,  
the salt effect  on the surface tension $\gamma$ has been 
examined extensively  for an 
air-water interface \cite{Jones,Levin-Flores,Levinhyd}, while  
it has not yet been well investigated for 
a liquid-liquid (oil-water) interface \cite{Nitro,Luo}. 
For a small amount of 
 solute,  most experimental interpretations 
have been based on the Gibbs formula \cite{Safran,Gibbs}
$\gamma=\gamma_0 -k_BT \Gamma$, where 
$\gamma_0$ is the surface tension without solute  and 
$\Gamma$ represents the  adsorption 
of solute per unit area. However, 
we have recently shown the presence of 
 a negative electrostatic contribution 
 $\gamma_{\rm el}$   for  a small amount of charged solute. 
 The  resultant generalized Gibbs formula 
is written as 
\cite{Current,OnukiJCP,OnukiEPL}, 
\be 
\gamma= \gamma_0 -k_BT \Gamma+\gamma_{\rm el}.  
\en 
Using 
the electric flux density 
 $\bi D$   
and  the electric field  ${\bi E}$, we have  
\bea 
\gamma_{\rm el}&=& - \int dz  {\bi D}\cdot{\bi E}/8\pi\nonumber\\
&=& - \int dz  {\rho}{\Phi}/2, 
\ena 
where the integral is in the surface normal direction ($\parallel z$).
In the second line 
we have used 
 the Poisson equation $\nabla\cdot{\bi D}= 4\pi\rho$  
and  the electric field  expression 
${\bi E}= -\nabla\Phi$,  
where $\rho$ is the charge density and 
$\Phi$ is the electric potential. Here,   
$\bi D$, $\bi E$, and $\rho$ are supposed to 
vanish  far from the interface. 
In this paper we also assume 
${\bi D}= \ve{\bi E}$, where 
$\ve$  is a  dielectric constant. 
The electrostatic contribution 
 is crucial for antagonistic salts \cite{OnukiJCP}  
and ionic surfactants \cite{OnukiEPL}. 
This result  has been derived 
from the calculation of the excess grand potential 
at a planar interface. In this paper, 
we will calculate 
the excess free energy  
for  small surface deformations, 
which is indeed expressed in terms of $\gamma$ in eq.(2).

In this paper, we also  discuss the surface instability 
induced by an antagonistic salt 
 consisting   of hydrophilic and hydrophobic ions 
 \cite{Araki,Nara,Current,Seto1,Seto2,Seto3}.   
An example is  sodium tetraphenylborate  NaBPh$_{4}$, which   
dissociates into hydrophilic Na$^+$ with $g_1 \gg 1$ 
and hydrophobic  BPh$_{4}^-$ with   $-g_2\gg 1$.  
The latter anion  consists  of  four phenyl rings bonded 
to an ionized boron. 
Such ion pairs   behave 
antagonistically in aqueous mixtures in the presence of composition 
heterogeneity. First,  they undergo 
microphase separation around a  liquid-liquid interface 
 on the scale of the Debye 
 screening length. This unique ion distribution produces    a large electric 
double layer 
\cite{OnukiPRE,OnukiJCP}, leading 
to a large   decrease in the surface tension \cite{OnukiJCP} 
in agreement with experiments \cite{Nitro,Luo}. 
From  x-ray   reflectivity measurements,  
Luo {\it et al.} \cite{Luo} determined 
 such ion distributions   
around a water-nitrobenzene(NB) interface 
by adding BPh$_{4}^-$ and two species of 
hydrophilic ions. 
Second, they  interact differently 
with water-rich and oil-rich 
composition fluctuations, leading to 
mesophases (charge density waves) near the solvent criticality.   
In accord with  this prediction,  Sadakane {\it et al.} 
\cite{Seto1} added a small amount of 
NaBPh$_{4}$     to a  near-critical 
mixture of D$_2$O and 3-methylpyridine (3MP) to find 
  a  peak   at an intermediate wave number 
$q_m$($\sim 0.1~$\AA$^{-1}\sim \kappa$) 
in the intensity  of small-angle neutron scattering. 
The peak height  was much 
enhanced with formation of periodic structures.    
(iii) Moreover, Sadakane {\it et al.}  observed  multi-lamellar (onion) 
structures   at  small volume fractions of 3MP 
(in D$_2$O-rich solvent)  far from 
the criticality \cite{Seto2}, 
where BPh$_{4}^-$  and solvating 3MP form charged lamellae. 
These  findings  demonstrate 
  very strong hydrophobicity of BPh$_{4}^-$. 
(iv) Another interesting phenomenon 
is    spontaneous emulsification 
(formation of small water droplets) 
at a water-NB interface \cite{Aoki,Poland}.
It was observed when a large pure water droplet was pushed  into 
a cell  containing NB and  antagonistic salt 
(tetraalkylammonium chloride). 

The organization of this paper is as follows. 
In Sec.2,   we will 
present the background 
of selective solvation. 
In Sec.3,  we will explain 
a   Ginzburg-Landau model 
for electrolytes   
accounting for  selective solvation.  
The surface tension formula (2) will be derived. 
In Sec.4, we will first calculate 
the free energy for a slightly deformed interface 
and then present simulation results on 
 the surface instability induced by antagonistic salt.

\section{Background of selective solvation }

Let us suppose  two species of ions ($i=1, 2$) 
with charges $Z_1e$ and $Z_2e$ ($Z_1>0$, $Z_2<0$). 
At  low ion densities, 
the  total ion chemical potentials $\mu_i$ 
in a  mixture solvent  are expressed as  
\be 
\mu_i= k_BT \ln (n_i\lambda_i^3) 
+ Z_ie\Phi+ \mu_{\rm sol}^i (\phi),
\en 
where $\phi$ is the water composition 
and $\lambda_i$ is the thermal de Broglie length 
(but is an irrelevant constant in the isothermal condition)  
and $\Phi$  is the local electric potential. 
For neutral hydrophobic particles 
the electrostatic term is nonexistent. 
The $\mu_i$  is a constant in equilibrium except for the 
interface region. 
We consider  a liquid-liquid  interface  between  
a  polar (water-rich) 
phase $\alpha$ and a less polar (oil-rich) 
phase $\beta$ with bulk compositions $\phi_\alpha$ 
and $\phi_\beta$ with $\phi_\alpha>\phi_\beta$.  
Thus   
$\mu_{\rm sol}^i(\phi)$ takes  different bulk values 
in the two phases due to its 
 composition dependence. So we define    
\be 
\Delta\mu_{\alpha\beta}^{i}
= \mu_{\rm sol}^{i}(\phi_\beta)- 
\mu_{\rm sol}^{i}(\phi_\alpha),  
\en 
which is called the Gibbs transfer free energy 
(per ion here) in electrochemistry
\cite{Current,Hung,Koryta,Sabela,Osakai,Ham}. 
The bulk ion densities far from the interface are written as 
$n_{i\alpha}$ in phase $\alpha$ and 
$n_{i\beta}$ in phase $\beta$.  From   the 
charge neutrality condition in  the bulk regions, 
 we  require 
$ 
Z_1n_{1\alpha}+Z_2n_{2\alpha}=0 
$ and 
$Z_1n_{1\beta}+Z_2n_{2\beta}=0.
$    
The potential    $\Phi$ tends to 
 constants  
$\Phi_\alpha$ and $\Phi_\beta$ in the  bulk 
 two phases, yielding   a  Galvani potential difference, 
\be  
 \Delta\Phi=\Phi_\alpha-\Phi_\beta,  
\en 
across the interface. 
Here  $\Phi$ approaches 
its limits  on the  scale of the Debye 
screening lengths,  $\kappa_\alpha^{-1}$ and 
$\kappa_\beta^{-1}$,  away from the interface, 
so we assume that 
the system  extends longer than 
$\kappa_\alpha^{-1}$ in phase $\alpha$ 
and $ \kappa_\beta^{-1}$ in phase $\beta$. 
For air-water interfaces,  there are  no 
ions in the air region, so $\kappa_\beta=0$, however.  
The continuity of $\mu_i$ across the interface  gives  
\be 
k_BT \ln (n_{i\alpha}/n_{i\beta})+ Z_i e\Delta\Phi 
- \Delta\mu_{\alpha\beta}^{i}=0,  
\en 
where $i=1,2$. After some calculations, 
$\Delta\Phi$ is expressed as 
 \cite{Hung,OnukiPRE}
\be
\Delta \Phi=[{\Delta\mu_{\alpha\beta}^1 
-\Delta\mu_{\alpha\beta}^{2}}]/{e(Z_1+|Z_2|)}. 
\en 
Similar potential differences also appear 
at liquid-solid interfaces (electrodes) \cite{Ham}. 
The ion densities in the bulk two phases (in the dilute limit) 
are simply related by  
\be 
 \frac{n_{1\beta}}{n_{1\alpha}}
= \frac{n_{2\beta}}{n_{2\alpha}} = 
\exp\bigg[- \frac{|Z_2| {\Delta\mu_{\alpha\beta}^{1}} +
Z_1 \Delta\mu_{\alpha\beta}^{2} }{(Z_1+|Z_2|)k_BT}\bigg] .  
\en 
However, if  three 
ion species are present,  the ion partitioning 
between two phases is much  more complicated 
\cite{OnukiJCP}.

The Gibbs transfer free energy  
has been determined  at present for  
water-nitrobenzene (NB)
\cite{Hung,Koryta,Sabela,Osakai}  
and water-1,2-dichloroethane(EDC) 
\cite{Sabela}  at room temperatures, where 
the dielectric constant of NB ($\sim 35$) 
is larger than that of  EDC ($\sim  10$).  
In the case of water-NB,  
the ratio  
  $\Delta\mu_{\alpha\beta}^i/k_{B}T$ 
  is $13.6$  for Na$^+$,   $27.1$  for Ca$^{2+}$, and 
$11.3$ for Br$^-$  as examples of 
hydrophilic ions, while it is $-14.4$ for  hydrophobic  
BPh$_4^-$. In the case of  water-EDC, 
it is  $22.7$  for Na$^+$ and   $17.5$  for Br$^-$, 
while it is $-14.1$ for    BPh$_4^-$.  
The amplitude  $|\Delta\mu_{\alpha\beta}^i|/k_{B}T$ for 
hydrophilic ions 
 is larger  for EDC than for NB and is very large for 
 multivalent ions.   
For  H$^+$ (more precisely  hydronium ions H$_3$O$^+$), 
 $\Delta\mu_{\alpha\beta}^i$   
 assumes  positive values close to those for  Na$^+$ 
 in  these two mixtures.

\section{Ginzburg-Landau theory of mixture electrolytes}
\subsection{Ginzburg-Landau free energy}

We consider  a  Ginzburg-Landau 
free energy $F$ for  a polar binary 
 mixture (water-oil)  
containing a small amount of a monovalent 
 salt ($Z_1=1$, $Z_2=-1$). 
  The ions  are  dilute and 
 their  volume fractions are  negligible.  
The variables  $\phi$, $n_1$, and $n_2$ are coarse-grained 
ones varying  smoothly on the molecular scale. 
We also  neglect the image 
interaction \cite{Levin-Flores}, 
though it  was included in our previous papers 
\cite{OnukiPRE,OnukiJCP}. 
See our previous analysis 
 \cite{OnukiPRE} 
 for  relative importance between the image interaction 
 and the solvation interaction at a liquid-liquid interface.
The  $F$ is the following space integral in the cell,
\be
F =\int d{\bi r}\bigg[f_{\rm tot} + \frac{1}{2}C |\nabla\phi|^2 
+ \frac{\varepsilon }{8\pi }{\bi E}^2 \bigg].  
\en
The first term $f_{\rm tot}$ depends  on 
$\phi$, $n_1$, and $n_2$ as   
\be
{f_{\rm tot}} = f(\phi) 
 + k_B T\sum_i n_i  \bigg[\ln (n_i\lambda_i^3) -1-  g_i \phi\bigg].
\en 
If the solvent molecular volumes of 
the two components  take 
a common value $v_0=a^3$ with $a\sim 3{\rm \AA}$, we may assume    
the    Bragg-Williams form 
\cite{Safran,Onukibook} for the first term,      
\be 
\frac{v_0f}{k_BT}   =   
 \phi \ln\phi + (1-\phi)\ln (1-\phi) 
+ \chi \phi (1-\phi),   
\en 
where   $\chi$ is the interaction 
parameter dependent on   $T$. 
The critical value of $\chi$ is 2 without ions. 
The $\lambda_i = \hbar(2\pi/m_i k_BT)^{1/2}$ in eq.(11) 
is   the thermal de Broglie wavelength of the species $i$ 
with $m_i$ being the molecular mass. 
The $g_1$ and $g_2$ are the solvation coupling 
constants assumed to be constants. 
The coefficient $C$  in the  gradient part of eq.(10) 
 is of order $k_BT/a$ 
 and should be determined  from the surface tension data 
or from the scattering data. 
The electric field is written as   ${\bi E}=-\nabla\Phi$. 
The  electric   potential $\Phi$ 
satisfies   the   Poisson equation,    
\be 
-\nabla\cdot\ve\nabla \Phi=  4\pi \rho. 
\en 
For simplicity, the dielectric constant $\ve$ is assumed to 
depend on $\phi$   as    
\be 
\ve(\phi)=\ve_0 + \ve_1 \phi. 
\en 
where $\ve_0$ and  $\ve_1$ are positive constants. 
A linear composition dependence of $\ve(\phi)$ 
was  observed by Debye and Kleboth 
for a mixture of 
nitrobenzene-2,2,4-trimethylpentane \cite{Debye}. 

The   solvation terms ($\propto g_i$)  
  follow 
if   $\mu_{\rm sol}^i(\phi)$ ($i=1,2$) 
 depend on $\phi$ linearly as  
\be 
\mu_{\rm sol}^i(\phi) =A_i   -k_B Tg_i\phi,    
\en 
where  $g_i$ are 
 constants independent of $\phi$. 
 We assume the presence of well-defined 
solvation shells in the concentarion 
range $\phi>\phi_{\rm sol}^i$ 
for hydrophilic ions (see the sentences below eq.(1)). 
The first term    $A_i$  is a constant  yielding    
 a  contribution linear with respect 
 to $n_i$ in $f_{\rm tot}$, so it  
 is irrelevant at constant ion numbers. 
The difference of the solvation 
chemical potentials  in two-phase coexistence  in eq.(5) 
 is given by  
$
\Delta\mu_{\rm sol}^i(\phi)= k_B Tg_i\Delta\phi,
$ 
where $\Delta\phi=\phi_\alpha-\phi_\beta$ is the 
composition difference. 
From eqs.(7) and (8), 
the Galvani potential difference  
and the  ion reduction factor are expressed in terms of 
$g_i\Delta\phi$ as  
\bea 
&&\Delta\Phi= k_BT (g_1-g_2)\Delta\phi /2e, \\
&&{n_{1\beta}}/{n_{1\alpha}}=
{n_{2\beta}}/{n_{2\alpha}}= 
\exp[- (g_1+g_2){\Delta\phi}/{2}].
\ena

The equilibrium interface profiles 
can be calculated by  requiring  the homogeneity of 
the chemical potentials $h= \delta F/\delta\phi$ and  
$\mu_i= \delta F/\delta n_i$. Some calculations give   
\bea 
&&h=f' -C\nabla^2\phi -\frac{\ve_1}{8\pi}{\bi E}^2 
- k_BT \sum_i g_i n_i,\\
&&\mu_i= k_BT [\ln(n_i\lambda_i^3)- g_i\phi]+ Z_ie\Phi ,
\ena 
where $f'=\p f/\p \phi$. The  ion distributions are 
expressed in terms of  $\phi$ and $\Phi$ 
in the modified Poisson-Boltzmann relations \cite{OnukiPRE},  
\be 
n_i=  n_i^0 \exp[g_i \phi -Z_i e\Phi/k_BT].
\en
The coefficients 
 $n_i^0$ are  determined 
from the conservation of the ion numbers, 
$\av{n_i}= V^{-1} 
\int d{\bi r}n_i({\bi r})=n_0,
$  
where $\av{\cdots}=V^{-1}\int d{\bi r}(\cdots)$ 
denotes the space average 
with $V$ being the cell volume. The average $n_0
=\av{n_1}=\av{n_2}$ 
is a given constant density.  

\subsection{Interface profiles 
and surface tension}

First, we give typical examples of 
the interface profiles for $\ve_1/\ve_0= 4/3$ and 
$e^2/\ve_1 k_BT= 20/\pi$.
In Fig. 1, 
we  display  the ion density $n_1(z)$  
and the  density difference $n_1(z)-n_2(z)$ 
proportional to the charge density for 
hydrophilic ion pairs with $g_1=4$ and $g_2=2$. 
We use the free energy density in eq.(12) 
and  set   $\chi=2, 1.95$, and 1.92. 
Note that the  critical value 
 of $\chi$  is 
 shifted as $\chi_c= 2+(g_1+g_2)^2v_0n_0/2$ 
 in the presence of ions in the mean-field theory.  
As $\chi \to \chi_c$, 
the  electric double layer at the interface 
diminishes. On the other hand, 
the ion distributions for  a pair of strongly  hydrophilic 
 and hydrophobic ions are very singular. 
In the left panel of Fig.2, 
we show  the ion densities 
for antagonistic ion pairs with $g_1=-g_2=10$,
 where $\chi=3$  and $v_0 n_{1\alpha}=v_0 n_{1\beta}= 
2\times 10^{-4}$. For this 
hydrophilic and hydrophobic ion pair, a 
 microphase separation 
forming  a  large electric double layer 
is apparent.

\begin{figure}
\begin{center}
\includegraphics[width=\textwidth, bb= 0 0 1109 443]{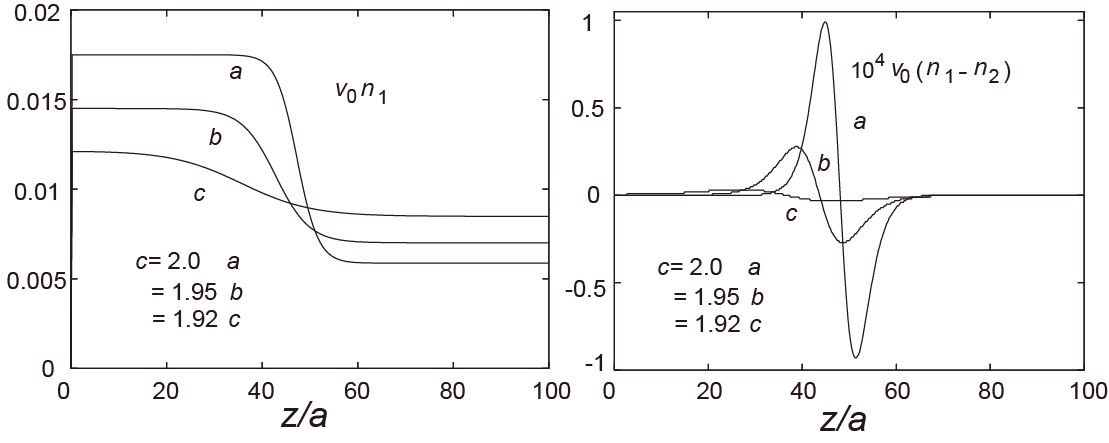} 
\end{center}
\caption{
Normalized ion  density $v_0n_1(z)$ (left) 
and normalized charge density 
$ v_0(n_1(z)-n_2(z)) $ (multiplied by $10^4$) (right) 
 with  $g_1=4$ and $g_2=2$ for 
 $\chi=2$ in (a), $1.95$ in (b), and 1.92 in (c). 
}
\end{figure}

\begin{figure}
\begin{center}
\includegraphics[width=\textwidth,bb= 0 0 1095 413]{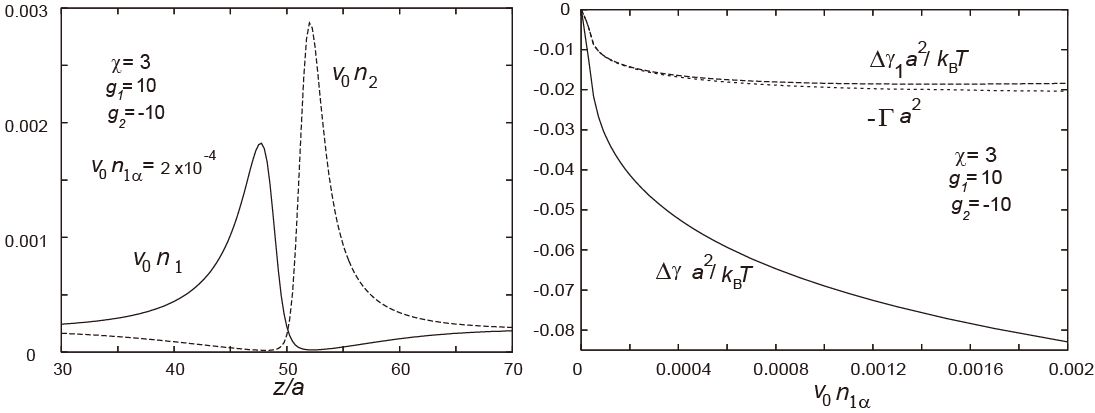} 
\end{center}
\caption{ 
Left: Normalized 
ion densities, where $\chi=3$,  
$g_1=-g_2=10$, and $v_0 n_{1\alpha}=v_0 n_{1\beta}= 
2\times 10^{-4}$. For this 
hydrophilic and hydrophobic ion pair, a 
 microphase separation 
forming  a  large electric double layer 
is apparent. 
Right: $a^2\Delta\gamma /k_BT$ 
 and $a^2\Delta\gamma_1 /k_BT$  as functions of 
 $v_0 n_{1\alpha}$, 
where  $\gamma_1$ is  the first term in eq.(24)  
and  $\Delta\gamma_1=\gamma_1-\gamma_{0}$ 
is very close to $-\Gamma a^2$. 
This shows  that the electrostatic part 
$\gamma_{\rm e}=\gamma-\gamma_1=-\int dz \ve E^2/8\pi$ 
dominates 
over $\Delta\gamma_{1}\cong -k_BT\Gamma $.
}
\end{figure}


Next, we consider the surface tension 
of a liquid-liquid interface 
in our Ginzburg-Landau scheme, where all 
the quantities depend on $z$. In equilibrium we  minimize  
 the grand potential $\Omega=\int d{z}\omega$, 
where  the grand potential density $\omega$  is written as  
\be  
\omega= f_{\rm tot}  + \frac{C}{2} 
 \phi'^2 
+ \frac{\varepsilon }{8\pi }{\bi E}^2  
- h\phi- \mu_1n_1 -\mu_2n_2.
\en 
Using  eqs.(17) and (18) we 
find $d(\omega+\rho \Phi)/dz= 2C\phi'\phi''$.  
Hereafter $\phi'=d\phi/dz$ and $\phi''=d^2\phi/dz^2$. 
Thus  $\omega(z)$ tends to 
a common constant $\omega_\infty$ as $z\to \pm\infty$ 
and   
\be 
\omega= C\phi'^2-\rho\Phi +\omega_\infty.  
\en  
The surface tension $\gamma=\int dz [\omega(z)-\omega_\infty]$ 
is then written  as \cite{OnukiPRE,OnukiJCP}  
\be 
\gamma =
  \int dz \bigg[ C\phi'^2- \frac{\ve }{4\pi}{\bi E}^2\bigg].
\en 
From eq.(20) we may also 
devise the following formula, 
\be
\gamma = \int dz \bigg[f(\phi) + \frac{C}{2}\phi'^2   
- k_{B}T n  - 
h \phi-C_\alpha   \bigg]  
-\int dz \frac{\ve}{8\pi}{\bi E}^2,
\en
where  $C_\alpha=f(\phi_\alpha) - k_{B}Tn_\alpha -
h \phi_\alpha$ and    
$
h=[ f(\phi_\alpha) -f(\phi_\beta)-k_BT(n_\alpha-n_\beta)]
/\Delta\phi
$
with $n_\alpha$ 
and $n_\beta$ being the bulk values 
of $n=n_1+n_2$. In accord with eq.(2), 
the right hand side of eq.(24)  may be 
expanded for small $n_0$  as 
\be   
\gamma= \gamma_0 -k_BT \Gamma -
\int dz\frac{\ve }{8\pi} {\bi E}^2+\cdots,  
\en  
 where 
$\gamma_0$ is the surface tension for $n_0=0$. 
In the second term,   
$\Gamma$ is the preferential adsorption 
at  the interface expressed as  
\be 
\Gamma= \int dz \bigg[ n-n_\alpha -\frac{\Delta n}{\Delta\phi}
 (\phi-\phi_\alpha)\bigg],
\en 
where $\Delta n=n_\alpha-n_\beta$. 
Thus the second term is the solute correction 
in the Gibbs formula. 
The third term is the electrostatic contribution 
growing  for antagonistic salt\cite{OnukiJCP,Araki}   and  
ionic surfactant \cite{OnukiEPL}.   
Thus eq.(25) is the generalized Gibbs formula 
including the electrostatic part.
In the right panel of  Fig.2, we examine  how 
 $\Delta\gamma=\gamma-\gamma_0$ 
and  $\Delta\gamma_1=\gamma_1-\gamma_0$
are  decreased with 
increasing $n_{1\alpha}$ for $g_1=-g_2=10$, 
where $\gamma_1$ is the first term on the 
right hand side of eq.(24). 
We notice the following. (i) The changes  
$\Delta\gamma$ 
and $\Delta\gamma_1$ are  
both proportional to  $n_{1\alpha}^{1/2}$ at small 
$v_0n_{1\alpha}$. 
Here $|\Delta\gamma|/(v_0n_{1\alpha})^{1/2}$ 
 is of order unity, so  $\Delta\gamma$    
is appreciable even for very small 
$v_0n_{1\alpha}$. This salt-density 
dependence was observed for a water-air interface 
\cite{Jones,OnukiJCP}.    
(ii) The electrostatic part  
$\gamma_{\rm e}\equiv -\int dz \ve E^2/8\pi$ 
is known to be important in this case 
from comparison between 
 $\Delta\gamma_1$ 
and $\Delta\gamma=\Delta\gamma_1+\gamma_{\rm e}$. 
 (iii) We confirm that the modified Gibbs relation 
$\Delta\gamma_1\cong -k_BT \Gamma$  holds excellently.

\section{Interface deformations}
\subsection{Free energy change}

We examine how the free energy changes  
for small deformations of the interface position 
 $z=\zeta(x,y)$. 
To this end, we superimpose  small deviations  
$\delta\phi$ and $\delta n_i$  on 
the equilibrium interface 
profiles,  $\phi=\phi(z)$ and $n_{i}=n_i(z)$, respectively. 
For simplicity, we replace $\phi(z)$ and $n_i(z)$  by 
 $\phi(z-\zeta)$ and $n_i(z-\zeta)$, respectively. 
 For small    $\zeta$. we set  
\be 
\delta\phi= -\phi'(z) \zeta, \quad 
\delta n_i= - n_i'(z)\zeta, 
\en   
where $\phi'= d\phi/dz$ and $n_i'= dn_i/dz$, 
Then the deviation of the dielectric constant 
 $\delta\ve= -\ve_1\phi'\zeta$ and 
 that of the charge density 
 $\delta\rho= -\rho'\delta\zeta$  are linear in $\zeta$, 
where $\rho'= e(n_1'-n_2')$. However, the electric potential  
is the solution of the Poisson equation (12) 
and its  deviation  
is expanded  as $\Phi_1+\Phi_2+O(\zeta^3)$ 
up to order $\zeta^2$, 
where $\Phi_1 \sim \zeta$ 
and $\Phi_2 \sim \zeta^2$. 
To first and second orders in $\zeta$, 
eq.(13) yields 
\bea 
&&\nabla\cdot[\delta\ve\nabla\Phi+ \ve\nabla\Phi_1]= 
4\pi \rho' \zeta,\\
&&\nabla\cdot[\delta\ve\nabla\Phi_1+ 
\ve\nabla\Phi_2]= 0. 
\ena  
Hereafter   $\Phi=\Phi(z)$ 
denotes the unpertubed potential for $\zeta=0$. 
The second-order change in the  electrostatic free energy 
$F_e=\int d{\bi r}\ve{\bi E}^2/8\pi$ is written as 
\bea 
\delta F_e&=& \int d{\bi r}\bigg[\frac{\ve}{8\pi} 
|{\nabla \Phi_1}|^2
+ \frac{\Phi'}{4\pi}\bigg
( \delta\ve \frac{\p\Phi_1}{\p z} 
+ \ve \frac{\p\Phi_2}{\p z} \bigg)\bigg] \nonumber\\
&=& \int d{\bi r}{\ve}
|{\nabla\Phi_1}|^2/{8\pi}.  
\ena 
If we multiply  eq.(29) by $\Phi$ and integrate over space,  
 we find $\int d{\bi r}{\Phi'}
( \delta\ve {\p\Phi_1}/{\p z} 
+ \ve {\p\Phi_2}/{\p z})=0$. 
Thus, in eq.(30), 
 the second line follows from the first line.  
Furthermore, taking the derivative of eq.(18) 
with respect to $z$ gives  
$
f''\phi'- C\phi'''- \ve_1\Phi'\Phi''/4\pi -k_BT\sum_i 
g_i n_i'=0
$, where $f''= \p^2 f/\p \phi^2$, 
$\phi'''= \p^3 \phi/\p z^3$, $\Phi'=d\Phi/dz$, 
and $\Phi''=d^2\Phi/dz^2$. We also find  
the relation $\sum_i (\delta n_i)^2/n_i=
\sum_i g_i\delta n_i \delta \phi +\delta\rho \Phi'\zeta$ 
from eq.(20).  
Using these relations, we calculate  
the second-order deviation of 
the free energy change $F$  as 
\be 
\delta F= 
 \int d{\bi r}\bigg[\frac{C}{2} \phi'^2|\nabla_\perp\zeta|^2 
 + \frac{\ve}{8\pi}
  \nabla \Phi_1 \cdot\nabla\varphi\bigg], 
\en 
where $\nabla_\perp= (\p/\p x,\p/\p y)$ 
and we introduce the combination  
$ \varphi= \Phi_1+\zeta\Phi'$ linear in $\zeta$. 
From eq.(28) $\varphi$ is related to $\zeta$ by 
\be 
\nabla \cdot\ve \nabla\varphi=  
\ve \Phi' \nabla_\perp^2\zeta,
\en 
so $\varphi=0$ for homogeneous $\zeta$. 
Using  the surface tension expression 
in eq.(23), we obtain    
\be 
\delta F=  \int d{\bi r}_\perp 
\frac{\gamma}{2} |\nabla_\perp \zeta|^2
+ \int d{\bi r}\frac{\ve}{8\pi}|\nabla\varphi|^2,
\en 
where $d{\bi r}_\perp= dxdy$ in the first term 
is the integral in the $xy$ plane. 
For homogeneous $\zeta$, we have  $\delta F=0$ 
since the deviations (26) 
represent a uniform translation of the interface. 

Let us consider the two-dimensional  
Fourier transform  
$\zeta_{\bi k}= \int d{\bi r}_\perp e^{-i{\bi k}\cdot{\bi r}_\perp}
\zeta({\bi r}_\perp)$, where ${\bi k}=(k_x,k_y)$ 
is the lateral wave vector. 
Then eq.(33) is rewritten as 
\be 
\delta F= \frac{1}{2}\int\frac{d{\bi k}}{(2\pi)^2} 
( {\gamma}k^2 + \Delta_k k^2) |{\zeta}_{\bi k}|^2
\en 
where $\Delta_k k^2$ arises from the second term in eq.(33). 
To calculate $\Delta_k$ the  
Fourier transform  
$\varphi_{\bi k}(z)= 
\int d{\bi r}_\perp e^{-i{\bi k}\cdot{\bi r}_\perp}
\varphi({\bi r}_\perp,z)$ is introduced.  From eq.(32)  
we may set $\varphi_{\bi k}(z)= -k^2
 G_{k}(z)\zeta_{\bi k}$, where $G_{k}(z)$
is a function of $z$ satisfying 
\be 
( \nabla_z \ve \nabla_z- \ve k^2)G_{k}
= \ve \Phi'.
\en 
In terms of $G_{k}$, $\Delta_k$ is 
expressed as 
\be 
\Delta_k= -k^2 \int dz~ \ve(z) \Phi'(z) G_k(z)/4\pi .
\en 
In particular, we consider the long wavelength limit 
with $k=|{\bi k}| \ll \xi^{-1}, \kappa_\alpha$, 
and $\kappa_\beta$, where $\xi$ is the interface thickness 
and $ \kappa_\alpha$  
and $\kappa_\beta$  are 
the Debye wave numbers in the bulk two phases. 
In the monovalent case they are given by 
\bea 
&&\kappa_\alpha= (8\pi n_\alpha e^2/\ve_\alpha k_BT)^{1/2}, \nonumber\\
&&\kappa_\beta= (8\pi n_\beta e^2/\ve_\beta k_BT)^{1/2}, 
\ena 
where $\ve_\alpha$ and $\ve_\beta$ are the dielectric 
constants in the two phases. In this case, 
for $|z|$ larger than $ \xi, \kappa_\alpha^{-1}$, 
and $\kappa_\beta^{-1}$, we have 
$G_k(z)=W_0  e^{-k|z|}/(\ve_\alpha+\ve_\beta)k$, 
where 
\be 
W_0= -\int dz~ \ve\Phi'.
\en  
Thus $\Delta_k$ is proportional to $k$ as    
\be 
\Delta_k= W_0^2k /4\pi(\ve_\alpha+\ve_\beta).  
\en 
If $\xi$ is shorter than 
$\kappa_\alpha^{-1}$ and  $\kappa_\beta^{-1}$, 
 the potential $\Phi(z)$ may be  calculated 
from  the nonlinear Poisson-Boltzmann 
equation\cite{OnukiJCP}, leading to 
\be 
W_0= \ve_\beta \Delta \Phi +\frac{k_BT}{e} (\ve_\alpha-\ve_\beta) 
\ln \bigg[\frac{1+b e^{\Delta U/2}}{1+b e^{-\Delta U/2}}\bigg],
\en   
where  $b=\ve_\beta\kappa_\beta/\ve_\alpha
\kappa_\alpha$ 
and  $\Delta U= e\Delta\Phi/k_BT$.
If  $\Delta U$ is small, 
we have 
$W_0 \cong (1+b)^{-1}( \ve_\beta +b \ve_\alpha)\Delta \Phi$, 
which simply  follows from the Debye-H$\ddot{\rm u}$ckel  approximation. 
In the simple case  $\ve_\alpha=\ve_\beta$, 
we have $W_0=  \ve_\beta \Delta \Phi$ and 
$\Delta_k= k_BT (\Delta U)^2 
\ell_B^{-1}  k^3/4\pi$, where 
$\ell_B= e^2/\ve_\alpha k_BT$ is the Bjerrum length common 
in the two phases. 

\subsection{Surface mode}

We consider the time development 
of the  surface perturbation $\zeta_{\bi k}(t)$. 
In the stable case, the linear damping rate 
 $\Omega_k$ is given by $\Omega_k= \gamma k/\eta$ 
 for $\Omega_k \ll \eta_0 k^2/\rho$, where 
   $\eta_0$ is the shear 
 viscosity\cite{Luca}.  Here we neglect 
 the acceleration term 
 in the momentum equation (as in eq.(44) below). 
  In our case with ions, some calculations  give 
\be 
 \Omega_k= (\gamma+\Delta_k) k/\eta_0. 
\en    
This result can also be used 
when  $\gamma<0$ is realized 
by a change in the temperature or in the salt amoun, 
where the surface 
perturbation grows in time at long wavelengths.

\subsection{Numerical results for surface instability 
induced by an antagonistic salt}

\begin{figure}
\begin{center}
\includegraphics[width=120mm, bb= 0 0 1230 939]{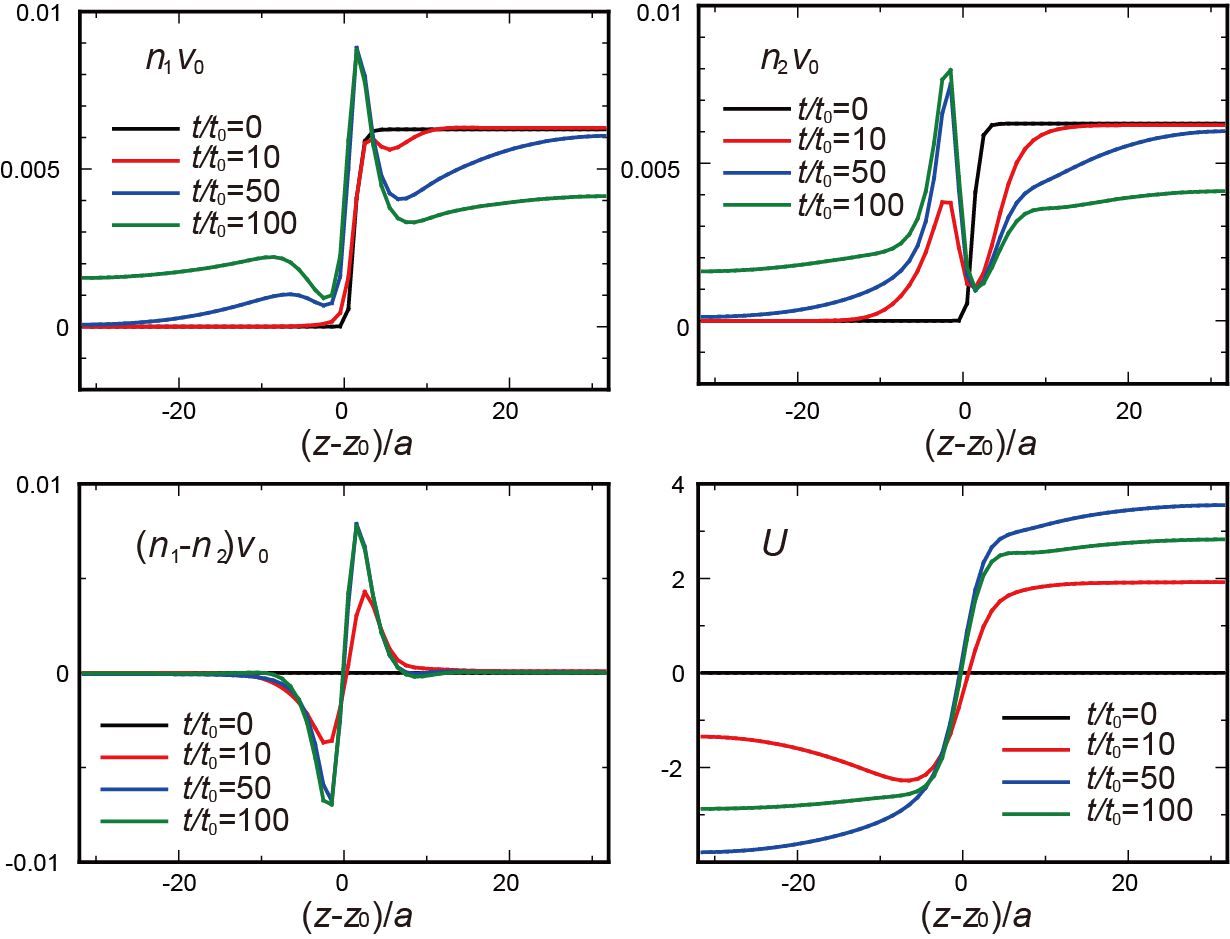}
\end{center}
\caption{(Color online) Time-evolution of 
$v_0n_1$ (left top),
$v_0n_2$ (right top),
$v_0(n_1-n_2)$ (left bottom),
and $U= e\Phi/k_BT$ (right top),
where a small amount of 
hydrophilic cations ($g_1=12$) and 
hydrophobic  anions ($g_2=-12$) 
are added in the water-rich region $z>0$ at $t=0$. 
An electric double layer is established 
on a time scale of $100t_0$, leading to a nnegative 
surface tension. 
}
\end{figure}

\begin{figure}
\begin{center}
\includegraphics[width=140mm, bb=0 0 396 476]{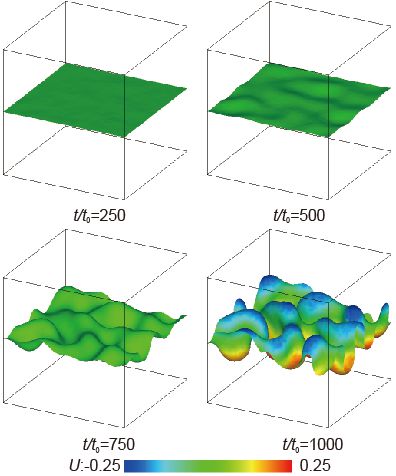}
\end{center}
\caption{(Color online) Slow growth of  interface 
deformations induced by 
antagonistic ion pairs 
for $t/t_0=250, 500,750$, and 1000, 
where the colors represent $U$ according to the color bar at 
the bottom. 
}
\end{figure}

As the right panel of Fig.2 indicates, the surface tension 
$\gamma$ can be made negative with addition of 
an antagonistic salt. Negativity of $\gamma$ 
can be realized particularly  near the solvent 
critical point even for small $n_0$, 
where the solvent surface tension  $\gamma_0$ 
becomes very small. 
When  $\gamma$ is slightly negative, 
we compare the two terms in eq.(34) to find 
a characteristic wave number in the early stage, 
\be 
k_c \sim  4\pi (\ve_\alpha+\ve_\beta) |\gamma|/W_0^2 ,    
\en 
where we use eqs.(33) and (38).
For $\ve_\alpha=\ve_\beta$, 
we have $k_c \sim  8\pi \ell_B 
|\gamma| /(k_BT |\Delta U|^2)$.
Near the solvent critical point, we have 
$\gamma_0 \sim 0.1 k_BT/\xi^2$ 
and  $k_c \sim  4\pi \ell_B 
|\gamma/\gamma_0| /(\xi\Delta\phi)^2(g_1-g_2)^2$.
We expect that  
the surface undulations  grow  
around  this wave number in the early stage.

 In this section,  we numerically examine 
the resultant interface  instability, 
where we initially  add  a small amount of 
 antagonistic ion pairs  with $  g_1=-g_2=12$ 
to the water side at $\av{n_1}=\av{n_2}= 
n_0 =0.003v_0^{-1}$.  
The initial interface position is at the middle of 
the cell, so $n_1(z,0)$ and $n_2(z,0)$ 
are equal to $2n_0$ above the interface 
and 0 below it  at $t=0$. 
The static parameters employed are  
$\chi=2.1$,  $C =k_BTa^{-1}$,  $\ve_1=0$, and 
$\ell_B= e^2/\ve_0 k_BT =3a$. 
The Debye wave number is then 
$\kappa=(8\pi n_0\ell_B)^{1/2}= 0.48 a^{-1}$.
Space is measured in units of $a= v_0^{1/3}$.

The  water composition $\phi$ 
and the ion densities $n_i$ obey \cite{Araki,Nara} 
\bea
&&\frac{\partial\phi}{\partial t}+
\nabla\cdot(\phi\mbox{\boldmath $v$})= 
\frac{L_0}{k_BT}  \nabla^2 h,\\
&&\frac{\partial n_i}{\partial t}+
\nabla\cdot(n_i \mbox{\boldmath $v$})= 
\frac{D_0}{k_BT}  \nabla \cdot n_i \nabla \mu_i .
\ena 
where $L_0$ 
is the kinetic coefficient, 
$D_0$ is the diffusion constant 
common to the cations and the anions, 
 and $h$ and $\mu_i$ are  defined by eqs.(17) and (18). 
Neglecting the acceleration term, we determine 
the velocity field $\bi v$ using  
the Stokes approximation,
\be 
\eta_0 \nabla^2{\bi v}
= \nabla p_1+ \phi \nabla h +\sum_i n_i\nabla \mu_i .
\en 
We introduce   $p_1$ to  ensure  
the incompressibility condition 
$\nabla\cdot{\bi v}=0$.  
We set $\eta_0L_0/k_BT=0.16a^4$ and $D_0= a^2/t_0$.
Time is  measured in units of  
\be 
t_0=a^5/L_0. 
\en 
The right hand side of eq.(45) is 
also written as $\nabla\cdot{\tensor \Pi}$, where  
${\tensor \Pi}$ is the stress tensor arising from  
the fluctuations of $\phi$ and $n_i$. 
Here  the total free energy $F$ in eq.(10) 
satisfies $dF/dt \le 0$ with these equations 
(if the boundary effect arising from 
the surface free energy is neglected).
We integrated eq.(43)  and (44) using 
 eq.(45)  at each step 
on a $64\times 64\times 64$ lattice 
under the periodic boundary condition in the $x$ and $y$ 
directions. The system length is then $L=64 a$. 
At the top $z=L/2$ and the bottom $z=-L/2$, 
we required  $E_z=- \p \Phi/\p z= 0$ 
and $\p n_i/\p z =0$.  
The system was initially in a two-phase 
with a planar interface at $z=0$ 
separating the two phases.

In Fig.3, we show the 
time development of 
$n_1$,  $n_2$, $n_1-n_2$ in the very early stage, 
and $\Phi$, which illustrate 
establishment of the electric double 
layer in the time region $t<100t_0$. 
In this stage, 
the profiles of $\phi$ and $n_i$ are nearly 
one-dimensional and  the 
surface undulations are negligible. 
We can see  establishment of a large electric double 
layer, 
where the  electric field is of order $k_BT/e$.  
The integral in eq.(23) is 
time-dependent here and may be regarded as an effective 
surface tension. It is about $0.2k_BT/a^2$ at $t=0$ 
and tends to a negative value of order 
$-0.05k_BT/a^2$ for $t\gs 50t_0$ in this case.  
Here,  $\Delta U \sim 5$ and  eq.(42) gives  $k_c\sim 
0.3 a^{-1}$.

In Fig.4, we display  
slower  time-evolution of  interface 
deformations after the establishment of the 
electric double layer  
induced by 
antagonistic ion pairs  
at $t/t_0=250, 500,750$, and 1000, 
where  the domain size is 
of order $2\pi/k_c$ in accord with 
eq.(42).   While the linear growth theory holds 
for $t\ls 500 t_0$, 
  the system is in the nonlinear growth regime 
with large  surface 
disturbances at $t=1000t_0$.  Note that this  final time 
is of the order of  the diffusion time $t_D= L^2/4D_0\sim 
10^3t_0$.  For $t \gg t_D$ 
the system should tend to a bicontinuous mesophase 
displayed in our previous paper \cite{Nara}.

\section{Summary}

We  stress   
the crucial role of the selective 
solvation of a solute in phase transitions of  various soft materials, 
which 
 should be relevant in  understanding a wide range  
of mysterious phenomena in water. 
Particularly remarkable  in polar binary mixtures are 
mesophase formation induced by  an antagonistic 
salt\cite{Current,Araki,Nara} 
 and precipitation induced by a  one-sided 
solute (a salt composed of 
hydrophilic cations and anions  and a  neutral 
hydrophobic solute)\cite{Okamoto,Opinion}.  
Regarding the  problem of mesophase 
formation, our theory 
is still insufficient and cannot 
well explain  the complicated  phase behavior 
disclosed by  the experiments 
\cite{Seto1,Seto2,Seto3}.

We propose experiments of 
the surface instability.   
For example, it should be induced  around a droplet 
of pure water (oil) inserted in 
a  bulk oil-rich  (water-rich) region 
in the immisible condition, where 
the droplet or the surrounding region  
contains  a small amount of an 
antagonistic salt. 
The  previous experiments 
of droplet formation\cite{Aoki,Poland}  
were  performed  in the strong segregation 
condition far from the solvent criticality 
and, as a result,   the surface tension might have  not 
been negative  even after the formation of 
an electric double layer. 
Thus, experiments should also be performed on  
near-critical aqueous mixtures in two-phase 
coexistence, where  
a small amount of an antagonistic salt 
is added to either of the water-rich 
or oil-rich region  as in Sec.4 of this paper.

\begin{acknowledgment}

This work was supported by 
KAKENHI (Grant-in-Aid for Scientific Research)
 from the Ministry of Education, Culture, 
Sports, Science and Technology of Japan. 
Thanks are due to informative discussions 
with   R. Okamoto, 
K. Sadakane, and H. Seto. 
\end{acknowledgment}

\end{document}